\documentclass[10pt,conference]{IEEEtran}
\IEEEoverridecommandlockouts
\usepackage{cite}
\usepackage{amsmath,amssymb,amsfonts}
\usepackage{algorithmic}
\usepackage{graphicx}
\usepackage{textcomp}
\usepackage{xcolor}

\usepackage{amsthm}
\usepackage{booktabs}
\usepackage{algorithm}
\usepackage{times}
\usepackage{soul}
\usepackage{booktabs}   
\usepackage{multirow}
\usepackage{adjustbox}
\usepackage{subcaption}
\usepackage{graphicx}
\usepackage{bm}
\usepackage{amssymb}
\usepackage{stfloats}

\def\BibTeX{{\rm B\kern-.05em{\sc i\kern-.025em b}\kern-.08em
    T\kern-.1667em\lower.7ex\hbox{E}\kern-.125emX}}
\begin{document}

\title{ControlMol: Adding Substructure Control To Molecule Diffusion Models \\
}

\author{\IEEEauthorblockN{Zhengyang Qi}
\IEEEauthorblockA{\textit{University of Science and } \\
\textit{Technology of China}\\
Hefei, China \\
miloq@mail.ustc.edu.cn}
\and
\IEEEauthorblockN{Zijing Liu}
\IEEEauthorblockA{\textit{International Digital Economy Academy} \\
\textit{(IDEA)}\\
Shenzhen, China \\
liuzijing@idea.edu.cn}
\and
\IEEEauthorblockN{Jiying Zhang}
\IEEEauthorblockA{\textit{International Digital Economy Academy} \\
\textit{(IDEA)}\\
Shenzhen, China \\
zhangjiying@idea.edu.cn}
\and
\IEEEauthorblockN{He Cao}
\IEEEauthorblockA{\textit{International Digital Economy Academy} \\
\textit{(IDEA)}\\
Shenzhen, China \\
caohe@idea.edu.cn}
\and
\IEEEauthorblockN{Xiaohua Xu* \thanks{*Corresponding author.}}
\IEEEauthorblockA{\textit{University of Science and } \\
\textit{Technology of China}\\
Hefei, China \\
xiaohuaxu@ustc.edu.cn}
\and
\IEEEauthorblockN{Yu Li* }
\IEEEauthorblockA{\textit{International Digital Economy Academy} \\
\textit{(IDEA)}\\
Shenzhen, China \\
liyu@idea.edu.cn}
}

\maketitle

\begin{abstract}
 
 Due to the vast design space of molecules, generating molecules conditioned on a specific sub-structure relevant to a particular function or therapeutic target is a crucial task in computer-aided drug design. Existing works mainly focus on specific tasks, such as linker design or scaffold hopping, each task requires training a model from scratch, and many well-pretrained \textit{De Novo} molecule
generation model parameters are not effectively utilized. To this end, we propose a two-stage training approach, consisting of condition learning and condition optimization. In the condition learning stage, we adopt the idea of ControlNet and design some meaningful adjustments to make the unconditional generative model learn sub-structure conditioned generation. In the condition optimization stage, by using human preference learning, we further enhance the stability and robustness of sub-structure control. In our experiments, only trained on randomly partitioned sub-structure data, the proposed method outperforms previous techniques by generating more valid and diverse molecules. Our method is easy to implement and can be quickly applied to various pre-trained molecule generation models. 
\end{abstract}

\begin{IEEEkeywords}
Molecular conformation generation, denoising diffusion probabilistic model, human preference learning
\end{IEEEkeywords}

\section{Introduction}
Molecule generation has become a fundamental task in drug design. A multitude of diffusion models~\cite{ddpm} automatically generating molecular geometries from scratch have been propose~\cite{edm,geodiff,torsion,midi,digress}. The primary goal of molecular design is to propose novel molecules that satisfy desired properties. However, they suffer from the huge space of diversity spaces. The space of pharmacologically-relevant molecules is estimated to exceed $10^{60}$  structures~\cite{virshup}. Searching in such space poses significant challenges for drug design. To reduce the size of the searching space, the strategy fragment-based drug design(FBDD)~\cite{fbdd} generates molecules from fragments~\cite{lead}. 
Some diffusion methods try to address this issue~\cite{diffhopp,difflinker,sbdd}, they improve the \textit{De Novo} molecule generation method by allowing it to receive 3D structure conditions.

EDM~\cite{edm} achieves equivariant unconditional molecular conformation generation. Difflinker~\cite{difflinker} extends the methods of EDM with a focus on linker design tasks, during the training process, it does not predict or update on fixed nodes. DiffSBDD~\cite{sbdd} further utilizes the methods of Difflinker to achieve generation conditioned over protein pockets, however, it only employs simple inpainting methods for tasks related to molecular structure control, such as Scaffold Hopping, Fragment Merging, and Fragment Growing.

Re-training a model for each task is resource-intensive, and the performance of training-free inpainting methods is not guaranteed (as will be demonstrated in our experiments). To address these issues, we propose ControlMol, a novel method that leverages the parameters of unconditional generative models to save training resources. Similar to the inpainting method, ControlMol is not limited by specific tasks.  Inspired by ControlNet~\cite{controlnet} and other similar works~\cite{T2I,gligen}, which adds visual modality's control to diffusion models in image generation, based on an unconditional generative model, we first utilize substructural data randomly partitioned from the complete molecule to train a sub-structure controlled generative model, we call this the condition learning stage.  Next, inspired by human feedback reinforcement learning (RLHF) in large language models and its applications in diffusion~\cite{dpo,diffusiondpo,diffusionrl}, we use DPO-like methods~\cite{dpo} to continue tuning the control model, enhancing its control capability, we call this the condition optimization stage.

Our contributions can be summarized as follows:
(1) We propose a new sub-structure controlled generation method distinct from all previous approaches in constrained molecule generation. It can generate more valid and diverse controlled molecules while saving training resources, and due to its excellent transferability, it can become a cornerstone for various molecule generation tasks in the future.
(2) We demonstrate the powerful effectiveness of reinforcement learning in this task and propose methods for stable reinforcement training.

\section{Method}

\begin{figure*}[t]
  \centering
  \includegraphics[width=1\linewidth]{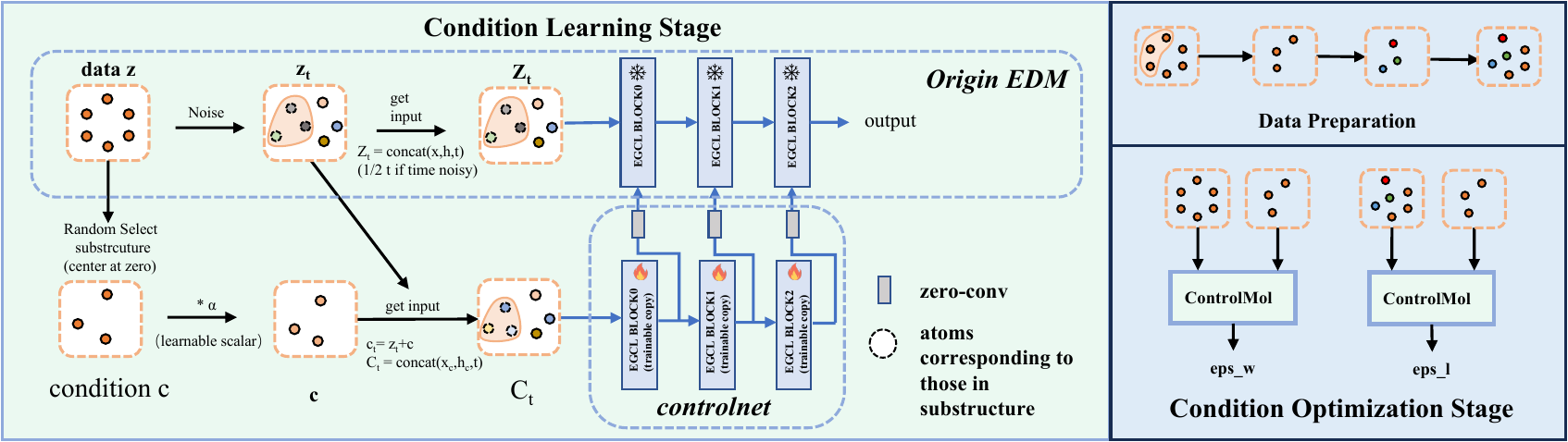}
  \caption{The overall architecture of ControlMol. Different colors of node represent different node features. After adding $c$, atoms corresponding to those in substructure in $C_t$ are more similar to the origin substructure both in terms of their position and node feature compared to $Z_t$, this will implicitly provide conditional information to the model.}
  \label{fig:net}
  \vspace{-10pt}
\end{figure*}

In this section, we introduce ControlMol, a unified sub-structure-conditioned diffusion model for molecules. Additionally, with properly setting, we can keep the E(3)-equivariant property of the base model. An overview of our method is presented in Figure \ref{fig:net}.

\subsection{Base Model}  \label{subsec:1}
\parskip=0pt
We use EDM\cite{edm} as our base model to introduce how to add 3D-substructure control to it.

\paragraph{Molecule Representation} 
We inherit the settings of EDM~\cite{edm}. A molecular with N atoms is represented as geometric graph $\mathcal G$ = $ \{\bm{x}, \bm{h}\}$ with coordinates $\bm{x}$ $ \in \mathbb{R}^{N \times 3}$ and node features $\bm{h}$ $\in \mathbb{R}^{N\times d}$.
\paragraph{Network Architecture}
Many works~\cite{edm,difflinker,sbdd} use E(n) Equivariant Graph Neural Networks(EGNNs)~\cite{egnn} as their model block. In EDM's setting,  EGNN consists of the composition of Equivariant Graph Convolutional Layers. $\bm{x}^{l+1}, \bm{h}^{l+1} = \mathrm{EGCL}[\bm{x}^l, \bm{h}^l]$ which are defined as:
\begin{align}
\mathrm{m}_{ij} &= \phi_{e}\left(\bm{h}_{i}^{l}, \bm{h}_{j}^{l}, d_{ij}^2\right), \, \bm{h}_{i}^{l+1} = \phi_{h}(\bm{h}_{i}^l, { \sum_{j \neq i}} \mathrm{m}_{ij}),  \nonumber \\
\bm{x}_{i}^{l+1} &= \bm{x}_{i}^{l}+\sum_{j \neq i} \frac{\bm{x}_{i}^{l}-\bm{x}_{j}^{l}}{d_{ij}+ 1} \phi_{x}\left(\bm{h}_{i}^{l}, \bm{h}_{j}^{l}, d_{ij}^2\right),
\label{eq:edm} 
\end{align}
where $l$ indexes the layer, $d_{ij} = \| x_{i}^{l}-x_{j}^{l} \|_2$ and $\phi_{e}$, $\phi_{h}$,$\phi_{x }$ are all learnable functions parameterized by fully connected neural networks.

At time t, EDM predicts noise $\hat{\bm{\epsilon}}$ includes coordinate and feature components: $\hat{\bm{\epsilon}}=[\hat{\bm{\epsilon}}^{x}, \hat{\bm{\epsilon}}^{h}]$. 
To make the network $\varphi$ invariant to translations, the initial coordinates from the coordinate component of the predicted noise are subtracted:
\begin{align}
    \hat{\bm{\epsilon}}=[\hat{\bm{\epsilon}}^x, \hat{\bm{\epsilon}}^h]=\varphi(\bm{z}_t,t)=\text{EGNN}(\bm{z}_t,t)-[\bm{z}_t^x, 0].
\end{align}

\subsection{Condition Learning Stage}  \label{subsec:2}

\paragraph{controlnet}
Unlike Unets~\cite{unet} in ControlNet~\cite{controlnet}, EGNN consists of EGCL blocks, there are no explicit residual connections between its layers. For this reason, as in Figure \ref{fig:net}, we choose to replicate all layers of the primary network and introduce control mechanisms at each layer. The EGCL update process in Equation \ref{eq:edm}, after getting the update by EGCL block, 
$\bm{x}_{i}^{l+1}, \bm{h}_{i}^{l+1}$ get extra update from controlnet\footnote{ControlNet corresponds to the paper~\cite{controlnet}, while controlnet denotes replicated blocks.} :
\begin{align}
\bm{x}_{i}^{l+1} &= \bm{x}_{i}^{l+1} +  \operatorname{zeroconv}(\Delta control\_\bm{x}_i^{l+1}) \nonumber \\ 
\bm{h}_{i}^{l+1} &= \bm{h}_{i}^{l+1} + \operatorname{zeroconv}(control\_\bm{h}_{i}^{l+1}) ,
\label{eq:controledm} 
\end{align}
where $control$ prefix represents the features in controlnet, and  $\Delta control\_\bm{x}_{i}^{l+1} =  control\_\bm{x}_{i}^{l+1}-control\_\bm{x}_{i}^{l} $ denotes the updated $\bm{x}$ in controlnet.

\paragraph{zero-conv} 
zero-conv in ControlNet~\cite{controlnet} is a $1\times1$ convolution layer with both weight and bias initialized with zeros. In our work, we adopt the name "zero-conv". Due to the different data types, we choose one-layer Linear with weight and bias initialized with zero and learnable scalar initialized zero as our zero-conv. Learnable scalar can keep the E(3)-equivariant property of EDM. However, by experimental comparison, we found that choosing Linear as the zero-conv can accelerate the convergence of the model and enhance the control effectiveness.  
\paragraph{Get Condition (sub-structure selection)} 
For each datapoint, we sample a proportion $\bm{p}\sim U(0,1)$, and each atom is selected in sub-structure with this probability. Then we set the center of mass of $\bm{u}$ to be zero. In particular, we do not employ an additional encoder for feature extraction and instead directly feed it into controlnet.  The input to controlnet at time $t$ is:
\begin{align}
    \bm{h} &= \bm{h}_{t} + \operatorname{zeroconv}(sub\_\bm{h}) \nonumber \\
    \bm{x} &= \bm{x}_{t} + \operatorname{zeroconv}(sub\_\bm{x})   
\label{eq:input}
\end{align}
where $sub$ prefix represents the features in 3D-context $\bm{u}$.

\paragraph{Time Noisy} By experimentation, we observed that employing the ControlNet methodology in a straightforward way to an unconditional model is hard to train (difficulty in convergence or slow convergence rate). We reason that a well-trained model can estimate the data distribution effectively by itself, it may tend to overlook the information from additional modules. We drop some time information to relax the module capacity. Specifically, In EDM's setting, it concats node feature $\bm{h}$ and time $t$ to add time information. At time $t$, we treat the input as follows (both in training and sampling): 
\begin{align}
    \bm{h} &= \operatorname{concat}(\bm{h}_\texttt{feature},t/2)  \ \text{(in main network)}  \nonumber \\
    \bm{h} &= \operatorname{concat}(\bm{h}_\texttt{feature},t)\ \ \ \      \ \text{(in controlnet)}
\end{align}

\begin{figure*}[t!]
\centering
\includegraphics[width=.99\textwidth]{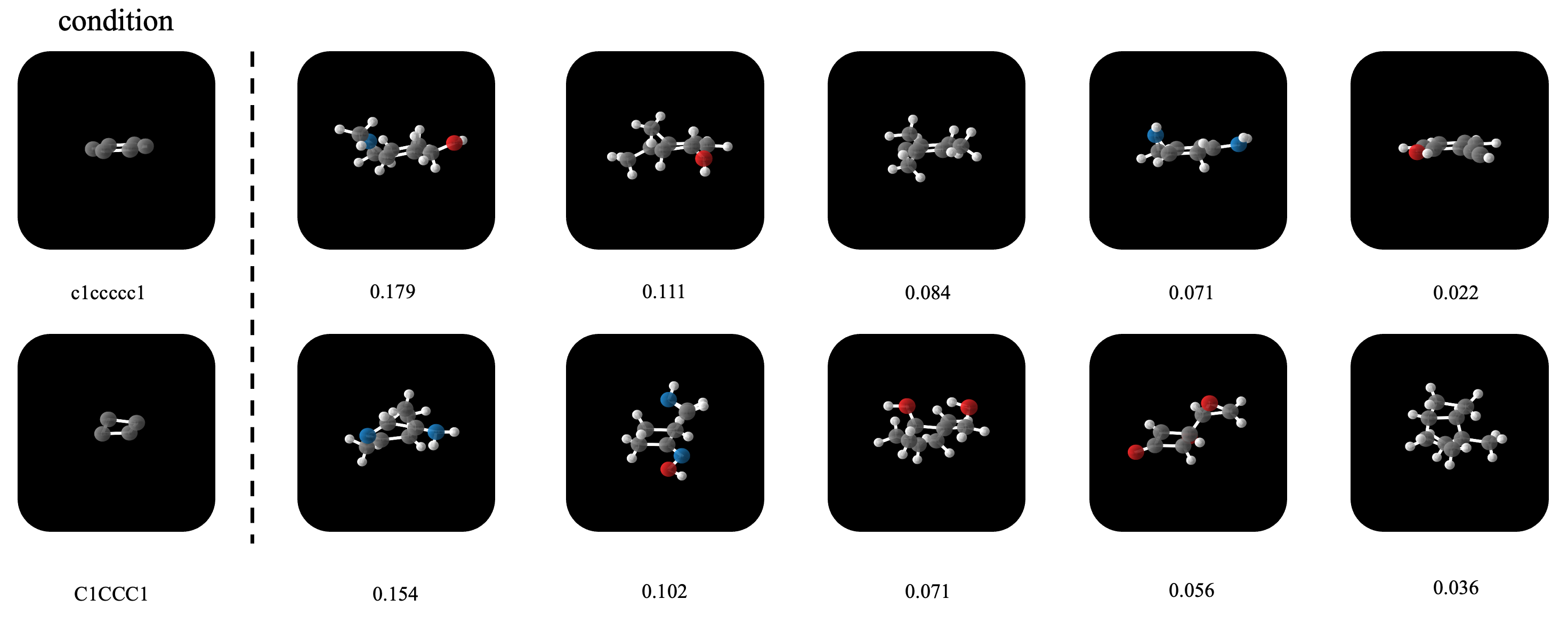}
\vspace{-10pt}
\caption{
Samples conditioned on "c1ccccc1" and "C1CCC1", we use RDkit to generate their conformer from smiles and sample condition on it. The number behind the figure is the \textbf{RMSD} between samples with the conditioned conformer (structure on the left). 
}
\label{fig:samples}
\vspace{-15pt}
\end{figure*}

\subsection{Condition Optimization Stage}  \label{subsec:3}
During the condition learning stage, due to the prevalence of C and H atoms in the dataset, the model's condition learning for other atoms is insufficient. Therefore, we designed the condition optimization phase specifically to reinforce the training of conditions. Similar to the condition learning stage, We freeze the parameters of the base model and only update the parameters of the controlnet.

\paragraph{Data Preparation} We consider a setting where we have a dataset $D=\{(\bm{c}, \bm{x_0}^w, \bm{x_0}^l)\}$ where each example contains a substructure $\bm{c}$ and a pairs of molecular with human label $\bm{x_0}^w \succ \bm{x_0}^l$. Same in the Condition learning Stage, for each datapoint $\bm{d}$, we sample a sub-structure $\bm{u}$. Next, we add noise to $\bm{u}$, while keeping the rest in $\bm{d}$ unchanged, then we obtain $\bm{d'}$. We consider $(\bm{u},\bm{d},\bm{d'})$ as the triplet data.

\paragraph{Training Objective}
We use the dpo loss function in DiffusionDPO~\cite{diffusiondpo}. 
\begin{multline}
    L_{dpo}(\theta)
    = - E_{
    (\bm{x_0}^w, \bm{x_0}^l) \sim D, t\sim \mathcal{U}(0,T), %
    \bm{x_t}^w \sim q(\bm{x_t}^w| \bm{x_0}), \bm{x_t}^l \sim q(\bm{x_t}^l| \bm{x_0})
    } \\
    \log\sigma (-\beta T ( \\
    \| \epsilon^w -\epsilon_\theta(\bm{x_t}^w,t)\|^2_2 - \|\epsilon^w - \epsilon_\text{ref}(\bm{x_t}^w,t)\|^2_2 \\
   - ( \| \epsilon^l -\epsilon_\theta(\bm{x_t}^l,t)\|^2_2 - \|\epsilon^l - \epsilon_\text{ref}(\bm{x_t}^l,t)\|^2_2)
    )\label{eq:loss-dpo-1}
\end{multline}

where $x_t^* =  \alpha_t x_0^*+\sigma_t\epsilon^*, \epsilon^* \sim N(0,I)$ is a draw from $q(x_t^*| x_0^*)$. Specifically, we only calculate the $\epsilon$ for the condition nodes, allowing the model to learn the generation of the condition nodes more accurately.

Additionally, we add a NLL loss used in Condition Learning Stage\footnote{For more details about the base loss in EDM, we refer the reader to~\cite{edm}.}. The aim is to prevent the model from training failing while reducing the probability of generating both good and bad samples~\cite{iterdpo}. The final loss function is as follows, where $\alpha$ is a hyperparameter.
\begin{align}
    L_{dpo+NLL}(\theta) = L_{dpo}(\theta) + \alpha L_{NLL}({\theta})
\label{eq:dponll}
\end{align}

\section{Experiment}

\setlength{\tabcolsep}{4pt}
\begin{table*}[t]
\centering

\scalebox{0.85}{
 {\fontsize{10}{12}\selectfont
\begin{tabular}{lc|cccc|cccc}\\\toprule 
\multirow{2}{*}{Methods} & \multirow{2}{*}{substructure} &  \multicolumn{4}{c}{QM9(w/o H)} & \multicolumn{4}{c}{QM9(with H)}\\
~ &   & $\%$ h\_success $\uparrow$ &  RMSD $\downarrow$ & $\%$ Valid $\uparrow$ & $\%$ Unique $\uparrow$ & $\%$ h\_success $\uparrow$ &  RMSD $\downarrow$ & $\%$ Valid $\uparrow$ &$\%$ Unique $\uparrow$\\\midrule

EDM-origin & - & - & - & 97.5 & 94.3 & - & - & 91.9 & 90.7     \\\midrule 
\multirow{3}{*}{Inpainting}  & C1CC1  &  100 &  0 &  6.8  &  100  & 100 & 0 & 1.2 & 100 \\    
& C1CCC1 & 100 & 0 & 7.7 & 97.4 &  100 & 0 &1 & 100\\ 
& c1ccccc1 & 100 & 0 & 0.9 & 11.1 & 100 & 0 & 0 & -  \\\midrule 
\multirow{3}{*}{Difflinker} &  C1CC1 & 100 & 0 & 78.9 & 83.3 & - & - & - & -  \\
& C1CCC1 & 100 & 0 & 79.3 & 67.2 & - & - & - & -   \\
& c1ccccc1 & 100 & 0 & 49.9 &0.2 & - & - & - & -  \\\midrule
\multirow{3}{*}{ControlMol (Stage1)} & C1CC1 & 94.2 & 0.164 & \textbf{93.6} & 99.4 & 90.2 & 0.111 & \textbf{74.5} & 100  \\
& C1CCC1 & 96.4 & 0.071 & \textbf{93.2} & 99.0 & 89.6 & 0.088 & \textbf{82.4} & 99.7\\
& c1ccccc1 & 97.9 & 0.099 & \textbf{94.9} & 75.5 & 75.6 & 0.085 & \textbf{75.9} & 95.3 \\ 
\bottomrule
\end{tabular}
}
}

\caption{ The comparison over 1000 molecules of ControlMol(Stage1) and other baseline models on each conditional generation task. Three specific cyclic structures are chosen to show comparison results. EDM-Origin is the unconditioned base model we use. "-" in Difflinker means that it can not converge. }
\label{tab:qm9}
\vspace{-8pt}
\end{table*}

\subsection{Experiment Setup}
\paragraph{Dataset} We consider QM9 dataset. QM9 contains up to 9 atoms (29 atoms including hydrogens) and we use the same train/val/test partitions in EDM, which consists of 100k/18k/13k samples respectively for each partition.

\paragraph{Metrics}   We measure the validity, and uniqueness of the samples. Same in the EDM~\cite{edm}, we don't report the novelty. Additionally, to evaluate position control effectiveness, we estimate the Root Mean Squared Deviation (RMSD) between the generated and real sub-structure  coordinates, and for node type control, we use h\_success(samples  which  have the right substructure node
 type / sample number) to measure the proportion of the generated samples that have the correct node type of the context.  

\paragraph{Hyperparameter} We tuned the $\beta$ in Equation \ref{eq:loss-dpo-1} in [0.1,0.5,5,10,2000] and  selected 5.  For $\alpha$ in Equation \ref{eq:dponll}, we tuned in [0.1-20] and selected 20. Batch size 256 and learning rate 1e-5 were used in both two training stages.  
\paragraph{Baselines} 

We compare our method with the inpainting-like method in DiffSBDD~\cite{sbdd}. We train Difflinker~\cite{difflinker} in randomly selected data(same as training ControlMol) instead of the fragment-linker data to be another baseline.   

\begin{figure}[t!]
\vspace{-10pt}
\centering
\includegraphics[width=.45\textwidth]{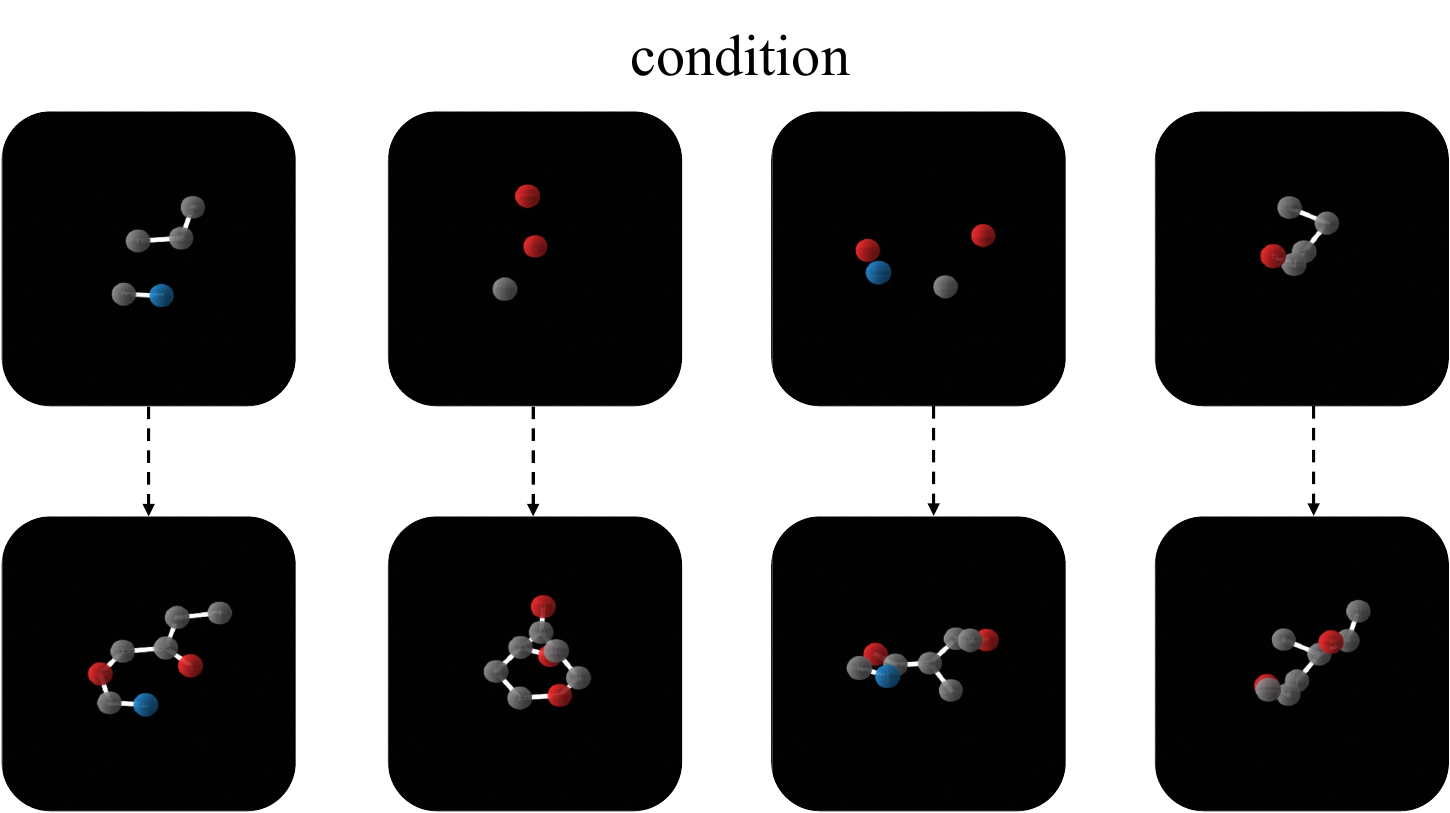}
\caption{
After the Condition Optimization Stage, ControlMol can achieve stable controlled generation on diverse atom types and structures.
}
\label{fig:diverse}
\vspace{-5pt}
\end{figure}

\setlength{\tabcolsep}{3pt}
\begin{table}[t!]
  \centering
   \begin{adjustbox}{scale=0.85}
   {\fontsize{10}{12}\selectfont
  \begin{tabular}{lcccc}
    \toprule
    Methods & $\%$ h\_success $\uparrow$ &  RMSD  $\downarrow$ & $\%$ Valid $\uparrow$ & $\%$ Unique $\uparrow$\\\midrule
    
    Stage1  & 33.4  &  0.581 &  97.3  &  99.62    \\  
                              
      Stage1+NLL  & 65.9  & 0.053  &  81.43  & 94.48  \\   
      
      Stage1+DPO & -  & - &  -  & -  \\

       Stage1+DPO+NLL  & \textbf{76.5}  &   \textbf{0.032} &   80.16  & 93.26  \\  
                   
        \bottomrule

  \end{tabular}
  }
  \end{adjustbox}
  \caption{The comparison over 10000 molecules(1000 for each smiles) of different loss in Stage2. "-" means the model can not converge only with dpo loss. }
  \label{tab:stage2}

\end{table}

\subsection{Result}
In Table \ref{tab:qm9}. We report the performances of all the models in terms of four metrics. All methods share the same architecture of EDM~\cite{edm}. As shown, ControlMol outperforms all the previous baseline methods. Specifically, we do not compare metrics with EDM-origin, even though we aim to maintain the capabilities of the base model as much as possible. However, since they are two different generation tasks, the metrics of EDM-origin and ControlMol cannot be compared straightforwardly.  

ControlMol generates all atoms while Difflinker only generates atoms excluding those in the sub-structure, so the structures in the samples may deviate slightly from those in the conditions and this deviation is assessed by RMSD. It's worth noting that the RMSD in Table \ref{tab:qm9} is the average of all samples, people can freely define a threshold to filter the desired molecules. In Figure \ref{fig:samples}, we show some samples with conditioned sub-structure and their RMSD.

Inpainting methods perform poorly in all cases. Difflinker can work in QM9 without H, but its performance sharply declines on "c1ccccc1". Difflinker even can't generate any valid molecules in QM9 with H. We believe it is due to the training data. "c1ccccc1" is rare in QM9 and due to randomly selecting sub-structure, it's hard for the model to see similar data in training data. The maximum number of atoms in QM9 is 29 when including Hydrogens, the data distribution with randomly selected substructures is more diverse, which makes the model hard to learn. Benefiting from the frozen parameters, ControlMol can work well in these cases, generating high ratio valid molecules.

In Table \ref{tab:stage2}, we select 10 substructures that contain non-carbon atoms(including CO, N=O, O=C=O). For each sub-structure, we sample 1000 samples and report the average metrics. As mentioned above, the sparse non-carbon data in the dataset can lead to insufficient learning by ControlMol for this portion. Consequently, ControlMol performs well on C/H structures, but its control effect on the other type atoms is less effective. After the reinforcement learning in Stage 2, ControlMol demonstrates robust control effects on diverse substructures. In Figure \ref{fig:diverse}, we show more samples after Stage2.

The reason why Stage1+NLL can enhance the results is that during Stage2 training, unlike in Stage1, we select all non-carbon atoms in the dataset as substructures for training, which increases the proportion of non-carbon data and enhances the learning effect.

\section{Conclusion}
We presented ControlMol, a method for adding 3D-structure control to diffusion models. ControlMol can generate more valid 3D-conditioned molecules and has more relaxed requirements for datasets compared to other methods. Our two-stage training method and data preparation can be leveraged for further work. Next, we will explore reinforcement learning training using molecules generated by ControlMol itself.

\end{document}